\documentstyle[12pt,epsf,a4]{article}

\newcommand{\beq}{\begin{equation}}
\newcommand{\eeq}{\end{equation}}
\newcommand{\beqa}{\begin{eqnarray}}
\newcommand{\eeqa}{\end{eqnarray}}
\newcommand{\no}{\nonumber}
\newcommand{\q}{\quad}
\newcommand{\qq}{\qquad}
\newcommand{\mnod}{\stackrel{\circ}{M}}

\begin{document}

\hfill 

\hfill 

\bigskip\bigskip

\begin{center}

{{\Large\bf The $\eta'$ and the topological charge density 
\footnote{Work supported in part by BMBF}}}

\end{center}

\vspace{.4in}

\begin{center}
{\large B. Borasoy\footnote{email: borasoy@physik.tu-muenchen.de}}

\bigskip

\bigskip

Physik Department\\
Technische Universit{\"a}t M{\"u}nchen\\
D-85747 Garching, Germany \\

\vspace{.2in}

\end{center}

\vspace{.7in}

\thispagestyle{empty} 

\begin{abstract}
We compare two different frameworks which have been proposed to include the
$\eta'$ in chiral perturbation theory. The equivalence of these two approaches
is shown both for the purely mesonic case and in the presence of the ground
state baryon octet.
The relation between the different sets of parameters in both Lagrangians is
clarified. 
\end{abstract}

\vfill

\section{Introduction}  
The QCD Lagrangian with massless quarks
exhibits an $SU(3)_R \times SU(3)_L$ chiral symmetry which is broken down
spontaneously to $SU(3)_V$, 
giving rise to a Goldstone boson octet of pseudoscalar mesons
which become massless in the chiral limit of zero quark masses.
On the other hand, the axial $U(1)$ symmetry of the QCD Lagrangian is broken by
the anomaly so that the mass of the corresponding pseudoscalar singlet does not
vanish in the chiral limit.
The lightest candidate would be the $\eta'$ with a mass of 958 MeV which is
considerably heavier than the octet states.
In conventional chiral perturbation theory the $\eta'$ 
is not included explicitly, although it does show up in the form of a
contribution to a coupling coefficient of the Lagrangian, a so-called
low-energy constant (LEC).

However, experiment suggests that the physical states --- $\eta$ 
and $\eta'$ --- are mixtures of octet and singlet components.
In order to include this effect in chiral perturbation theory one should treat
the $\eta'$ as a dynamical field variable instead of integrating it out from
the effective theory.
This approach is also motivated by large $N_c$ considerations. In this limit
the axial anomaly is supressed by powers of 1/$N_c$ and gives rise to a ninth 
Goldstone boson, the $\eta'$.
The inclusion of the $\eta'$ in chiral perturbation theory has been the subject
of previous work, see e.g. \cite{GL, Leu, RST}.
But while the authors of \cite{GL, Leu} work with a 
$U(3)_R \times U(3)_L$ invariant
Lagrangian, gluonic terms have been included explicitly in the effective
theory in \cite{RST}. 
The equivalence of both approaches is rather evident to lowest order both in
the chiral and the 1/$N_c$ expansion, e.g., the Lagrangian in Eq. (2.22)
of Di Vecchia's and Veneziano's work \cite{RST} coincides with the
corresponding part of Eq. (2) in Leutwyler's presentation \cite{Leu}.
But so far no systematic comparison between both
schemes has been made to prove the equivalence
at higher orders. The purpose of the present work is to fill this gap.

In the next two sections we will compare the mesonic Lagrangians in both
approaches and, furthermore,
generalize the approach of \cite{RST} to higher orders in
the gluonic terms.
Having shown the equivalence of both frameworks in the mesonic sector, we
proceed by including the ground state baryon octet in Section 4. The inclusion
of the $\eta'$ in baryon chiral perturbation theory has been the subject
of recent work \cite{B, Ste} and again the two different approaches have been
used without clarifying the connection between both schemes.
It is therefore desirable to show the equivalence also in the baryonic case.
We conclude with a short summary.

\section{The $U(1)_A$ invariant effective Lagrangian}
In this section we will briefly outline the method of extending the
$SU(3)_R \times SU(3)_L$ symmetry of the effective Lagrangian in conventional
chiral perturbation theory to $U(3)_R \times U(3)_L$ in a more generalized
framework including the $\eta'$, see e.g. \cite{GL,Leu}.
Within this approach 
the topological charge operator coupled to an external field is added to the
QCD Lagrangian
\beq  \label{lag}
{\cal L} = {\cal L}_{QCD} - \frac{g^2}{16 \pi^2} \theta(x) \mbox{tr}_c
  ( G_{\mu \nu} \tilde{G}^{\mu \nu} )
\eeq
with $\tilde{G}_{\mu \nu} = \epsilon_{\mu \nu \alpha \beta} G^{\alpha
\beta}$ and $\mbox{tr}_c$ is the trace over the color indices.
Under $U(1)_R \times U(1)_L$ the axial $U(1)$ anomaly 
adds a term $ -( g^2 / 16 \pi^2)
2 N_f \, \alpha \, \mbox{tr}_c ( G_{\mu \nu} \tilde{G}^{\mu \nu} )$ to the
QCD Lagrangian, with $N_f$ being the number of different quark flavors and
$\alpha$ the angle of the global axial $U(1)$ rotation.
The vacuum angle $\theta(x)$ is in this context treated as an external field
that transforms under an axial $U(1)$ rotation as
\beq  \label{theta}
\theta(x) \rightarrow  \theta'(x) = \theta(x) - 2 N_f \alpha .
\eeq
Then the term generated by the anomaly in the fermion determinant is
compensated by the shift in the $\theta$ source and the Lagrangian from
Eq. (\ref{lag}) remains invariant under axial $U(1)$ transformations.
The symmetry group $SU(3)_R \times SU(3)_L$ of the Lagrangian ${\cal L}_{QCD}$
is extended to $U(3)_R \times U(3)_L$ for ${\cal L}$.
\footnote{Note that the Lagrangian actually changes by a total derivative 
which gives rise to the Wess-Zumino term. We will disregard this 
contribution, since it is irrelevant for proving the equivalence of 
both schemes discussed in this presentation.}
This property remains at the level of an effective theory and the
additional source $\theta$ also shows up in the effective Lagrangian.
Let us consider the purely mesonic effective theory first.
The lowest lying pseudoscalar meson nonet is summarized in a matrix valued 
field $U(x)$
\begin{equation}
 U(\phi,\eta_0) = u^2 (\phi,\eta_0) = 
\exp \lbrace 2 i \phi / F_\pi + i \sqrt{\frac{2}{3}} \eta_0/ F_0 \rbrace  
, 
\end{equation}
where $F_\pi \simeq 92.4$ MeV is the pion decay constant and the singlet
$\eta_0$ couples to the singlet axial current with strength $F_0$.
The unimodular part of the field $U(x)$ contains the degrees of freedom of
the Goldstone boson octet $\phi$
\begin{eqnarray}
 \phi =  \frac{1}{\sqrt{2}}  \left(
\matrix { {1\over \sqrt 2} \pi^0 + {1 \over \sqrt 6} \eta_8
&\pi^+ &K^+ \nonumber \\
\pi^-
        & -{1\over \sqrt 2} \pi^0 + {1 \over \sqrt 6} \eta_8 & K^0
        \nonumber \\
K^-
        &  \bar{K^0}&- {2 \over \sqrt 6} \eta_8  \nonumber \\} 
\!\!\!\!\!\!\!\!\!\!\!\!\!\!\! \right) \, \, \, \, \, ,  
\end{eqnarray}
while the phase det$U(x)=e^{i\sqrt{6}\eta_0/F_0}$
describes the $\eta_0$ .
The symmetry $U(3)_R \times U(3)_L$ does not have a dimension-nine
irreducible representation and consequently does not exhibit a nonet symmetry. 
We have therefore used the different notation $F_0$ for the decay constant 
of the singlet field.
The effective Lagrangian is formed with the fields $U(x)$,  
derivatives thereof and also includes both the quark mass matrix ${\cal M}$ and
the vacuum angle $\theta$: ${\cal L}_{\mbox{eff}}(U,\partial U,\ldots,
{\cal M},\theta)$. Under $U(3)_R \times U(3)_L$ the fields transform as
follows
\beq
U' = RUL^{\dagger} \q , \qq {\cal M}'= R{\cal M}L^{\dagger} \q , \qq
\theta'(x) = \theta(x) - 2 N_f \alpha
\eeq
with $R \in U(3)_R$, $L \in U(3)_L$,
but the Lagrangian remains invariant. 
The phase of the determinant
det$U(x)=e^{i\sqrt{6}\eta_0/F_0}$ transforms  under axial $U(1)$ as
$\sqrt{6} \eta_0'/F_0 = \sqrt{6} \eta_0/F_0  + 2 N_f \alpha$ so that
the combination $\sqrt{6} \eta_0/F_0 + \theta$ remains invariant.
It is more convenient to replace the variable $\theta$ by this invariant
combination, ${\cal L}_{\mbox{eff}}= {\cal L}_{\mbox{eff}}(U,\partial U,\ldots,
{\cal M},\sqrt{6} \eta_0/F_0 + \theta)$.
One can now construct the effective Lagrangian in these
fields that respects the symmetries of the underlying theory.
In particular, the Lagrangian is invariant under $U(3)_R \times U(3)_L$
rotations of $U$ and ${\cal M}$ at a fixed value of the last argument.
The most general Lagrangian up to and including terms with two derivatives and
one factor of ${\cal M}$ reads
\beqa  \label{mes}
{\cal L}_{\phi} &=& - V_0 + V_1 \langle \nabla_{\mu}U^{\dagger} \nabla^{\mu}U 
\rangle  + V_2 \langle \chi_+ \rangle + i V_3 \langle \chi_- \rangle \no \\
&&+ V_4 \langle U^{\dagger} \nabla_{\mu}U  \rangle 
\langle U^{\dagger} \nabla^{\mu}U 
\rangle + i V_5\langle U^{\dagger} \nabla_{\mu}U \rangle \nabla^{\mu} \theta
+ V_6 \nabla_{\mu} \theta \nabla^{\mu} \theta .
\eeqa
The expression $\langle \ldots \rangle$ denotes the trace in flavor space
and the quark mass matrix ${\cal M} = \mbox{diag}(m_u,m_d,m_s)$
enters in the combinations
\beq
\chi_\pm = 2 B_0 ( u {\cal M} u \pm  u^\dagger {\cal M} u^\dagger)
\eeq
with $B_0 = - \langle  0 | \bar{q} q | 0\rangle/ F_\pi^2$ the order
parameter of the spontaneous symmetry violation.
The covariant derivatives are defined by
\beqa
\nabla_{\mu} U  &=&  \partial_{\mu} U - i ( v_{\mu} + a_{\mu}) U
                     + i U ( v_{\mu} - a_{\mu})   \no \\
\nabla_{\mu} \theta  & = &  \partial_{\mu} \theta + 2 \langle a_{\mu} \rangle .
\eeqa
The external fields $v_{\mu}(x),a_{\mu}(x)$ represent hermitian $3 \times 3$
matrices in flavor space.
Note that the term of the type $i \langle  U^{\dagger} \nabla_{\mu}U
\rangle \nabla^{\mu} \theta$
can be transformed away \cite{Leu}, but for our purposes it is  more convenient
to keep this term explicitly. 
Once the equivalence of both approaches is shown,
one is free to eliminate such a term.
The coefficients $V_i$ are functions of the variable 
$\sqrt{6} \eta_0/F_0 + \theta$, $V_i(\sqrt{6} \eta_0/F_0 + \theta)$,
and can be expanded in terms of this variable. At a given order of
derivatives of the meson fields $U$ and insertions of the quark mass matrix 
${\cal M}$ one obtains an infinite string of increasing powers of the 
singlet field $\eta_0$ with couplings which are not fixed by chiral symmetry.
Parity conservation implies that the $V_i$ are all even functions
of $\sqrt{6} \eta_0/F_0 + \theta$ except $V_3$, which is odd, and
$V_1(0) = V_2(0) = F_\pi^2/4$ gives the correct  normalizaton
for the quadratic terms of the Goldstone boson octet.

\section{The topological charge density within an effective Lagrangian}
In the literature, another approach of incorporating the axial $U(1)$ anomaly 
in an effective Lagrangian can be found \cite{RST}.
But so far no attempt has been made to compare this scheme with the approach
presented in the last section. 
In this section we will set up an effective Lagrangian following
the ideas of \cite{RST} and compare it with the Lagrangian from
Eq. (\ref{mes}).
The starting point is the effective Lagrangian 
\beq
{\cal L}_{\phi} =  \frac{F_\pi^2}{4} \langle \nabla_{\mu} U^{\dagger} 
     \nabla^{\mu} U \rangle  +  \frac{F_\pi^2}{4} \langle \chi_+ \rangle +
   a \langle U^{\dagger} \nabla_{\mu}U \rangle 
  \langle U^{\dagger} \nabla^{\mu}U \rangle 
\eeq
which reduces to conventional $SU(3)_R \times SU(3)_L$ chiral perturbation
theory if the singlet field $\eta_0$ is neglected.
A new low-energy constant $a$ enters the calculation which for our purposes
here will be left undetermined.
Next, one introduces gluonic terms in order to reproduce the anomaly in the
divergence of the axial-vector current
\beq
\partial_{\mu} J^{\mu}_5 =   2 i \sum_f m_f \bar{q}_f \gamma_5 q_f +
 \frac{g^2}{16 \pi^2} 2 N_f \mbox{tr}_c ( G_{\mu \nu} \tilde{G}^{\mu \nu} )
\eeq
by defining $Q(x) \equiv  (g^2/16 \pi^2) \mbox{tr}_c 
( G_{\mu \nu} \tilde{G}^{\mu \nu} ) $.
The correct transformation under axial $U(1)$ is achieved by adding the term
\beq
\delta {\cal L} = \frac{i}{2} Q \langle \log U - \log U^\dagger \rangle 
\eeq
to the effective Lagrangian, where it is assumed that the topological charge
density $Q(x)$ remains invariant under $U(1)_A$ transformations.
The most general effective Lagrangian in this framework up to and including
terms with two derivatives, one factor of $\cal{M}$ and quadratic terms in $Q$
respecting the symmetries of the underlying theory reads 
\beqa \label{q}
{\cal L}_{\phi} &=&  \Big( \frac{F_\pi^2}{4} + v_1 Q^2 \Big)
            \langle \nabla_{\mu} U^{\dagger} \nabla^{\mu} U \rangle  
  +  \Big( \frac{F_\pi^2}{4} + v_2 Q^2 \Big) \langle \chi_+ \rangle 
  + \kappa Q \no \\
  && + \frac{i}{2} Q \langle \log U - \log U^\dagger \rangle + \tau Q^2
  + i v_3 Q \langle \chi_- \rangle + 
   v_6 \partial_{\mu} Q \partial^{\mu} Q   \no \\
  && + \Big( a + v_4 Q^2 \Big)\langle U^{\dagger} \nabla_{\mu}U \rangle 
   \langle U^{\dagger} \nabla^{\mu}U \rangle 
  + i v_5 \langle U^{\dagger} \nabla_{\mu}U \rangle  \partial^{\mu} Q  
\eeqa
where an irrelevant constant has been omitted.
From matching to QCD we know that the parity-violating piece $\kappa Q $ of
the effective Lagrangian equals
\beq
\delta {\cal L} = - \theta \frac{g^2}{16 \pi^2} \mbox{tr}_c
  ( G_{\mu \nu} \tilde{G}^{\mu \nu} ) = - \theta Q . 
\eeq
We will therefore set $\kappa = - \theta$ in the following.
Usually authors have neglected some of these terms using only the Lagrangian
\beqa
{\cal L}'_{\phi} &=&  \frac{F_\pi^2}{4} \langle \nabla_{\mu} U^{\dagger} 
     \nabla^{\mu} U \rangle  +  \frac{F_\pi^2}{4} \langle \chi_+ \rangle +
   a \langle U^{\dagger} \nabla_{\mu}U \rangle 
  \langle U^{\dagger} \nabla^{\mu}U \rangle  \no \\
&&  + \frac{i}{2} Q \langle \log U - \log U^\dagger \rangle - \theta Q
 + \tau Q^2
\eeqa
in which $Q$ decouples from the Goldstone boson octet $\phi$.
This is motivated by the fact that the topological charge density $Q$ behaves
in the large $N_c$ limit as $Q \propto g^2 \propto 1/N_c$ and higher orders of
$Q$ are suppressed by powers of $1/N_c$.
In order to prove the equivalence of this approach to that of the last section,
we prefer to work with the Lagrangian in Eq. (\ref{q}).
The generalization of the proof to higher orders, both in the derivative
expansion and in $Q$, is straightforward and will be discussed later.
We will therefore restrict ourselves to this Lagrangian in the beginning.
The gluonic term $Q$ is treated as a background field and is integrated 
out from the Lagrangian via its equation of motion
\beq
\partial_{\mu} \frac{\delta {\cal L}}{\delta \partial_{\mu} Q} -
\frac{\delta {\cal L}}{\delta Q} = 0 .
\eeq
To lowest order in the derivatives and the quark masses the equation of motion
for $Q$ reads
\beqa  \label{low}
Q &=& \frac{1}{2 \tau} \Big( \theta - \frac{i}{2} \langle \log U - 
  \log U^\dagger \rangle \Big) \no \\
&=&  \frac{1}{2 \tau} \Big( \theta + \sqrt{6} \eta_0/F_0 \Big) \equiv 
\frac{1}{2 \tau} Q_0 .
\eeqa
Under axial $U(1)$ tansformations the $\eta_0$ field transforms as
$\sqrt{6} \eta_0/F_0 \rightarrow \sqrt{6} \eta_0/F_0  + 2 N_f \alpha$,
where $\alpha$ is the angle of the axial $U(1)$ rotation.
For $Q$ to remain invariant, $\theta$ has to compensate for the change in
$\eta_0$, cf. Eq. (\ref{theta}),
\beq  \label{the}
\theta \rightarrow   \theta - 2 N_f \alpha .
\eeq
It is therefore more convenient to consider $\theta$ as an external field
$\theta(x)$ which has under $U(1)_A$ the transformation property given in 
Eq. (\ref{the}) rather than to treat it as a constant (see the work by Di
Vecchia and Veneziano \cite{RST} for the latter case).
This leads to an effective Lagrangian which remains invariant also under 
$U(1)_A$ rotations in agreement with the first approach.
Otherwise, $Q$ would not be $U(1)_A$ invariant
in contradiction to the assumption.
Reinserting the solution $\frac{1}{2 \tau} Q_0$ of the equation of motion for
$Q$ into the Lagrangian in Eq. (\ref{q}) one obtains
\beqa  \label{high}
{\cal L}_{\phi} &=&  \Big( \frac{F_\pi^2}{4} + \frac{v_1}{4 \tau^2} Q_0^2 \Big)
            \langle \nabla_{\mu} U^{\dagger} \nabla^{\mu} U \rangle  
  +  \Big( \frac{F_\pi^2}{4} + \frac{v_2}{4 \tau^2} Q_0^2 \Big) \langle \chi_+
  \rangle - \frac{1}{4 \tau} Q_0^2  \no \\
  &&  + i \frac{v_3}{2 \tau}  Q_0 \langle \chi_- \rangle 
   + \Big( a + \frac{v_5}{2 \tau} - \frac{v_6}{4 \tau^2}+
   \frac{v_4}{4 \tau^2} Q_0^2 \Big)\langle U^{\dagger} \nabla_{\mu}U \rangle 
   \langle U^{\dagger} \nabla^{\mu}U \rangle  \no \\
 &&  + i  \Big( \frac{v_5}{2 \tau} - \frac{v_6}{2 \tau^2} \Big)
    \langle U^{\dagger} \nabla_{\mu}U \rangle \nabla^{\mu} \theta
  + \frac{v_6}{4 \tau^2} \nabla_{\mu} \theta \nabla^{\mu} \theta .
\eeqa
This Lagrangian is in complete agreement with the one in Eq. (\ref{mes}),
once one expands the functions $V_i$ in powers of 
$\sqrt{6} \eta_0/F_0 + \theta = Q_0$ and keeps only the first terms in the
expansions. There is a one-to-one correspondence between the low-energy
constants in both schemes to the order we are working.
This equivalence is maintained at higher orders both in the derivative
expansion and in the background field $Q$.

Firstly, we will examine the latter case by adding a piece $\delta {\cal L} =
\lambda Q^4$ to the Lagrangian. Other terms with higher powers of $Q$ can be
included in the Lagrangian as well, but they do not alter the following
considerations. In order to keep the presentation lucid, we restrict
ourselves to this simple extension.
The modified equation of motion for $Q$ reads to leading order in the
derivatives and quark masses
\beq  \label{eoml}
- Q_0 + 2 \tau Q + 4 \lambda Q^3 = 0 .
\eeq
Although this equation can still be solved analytically, we prefer to solve it
perturbatively, since this method can be generalized to arbitrary high powers
in $Q$.
The $1/N_c$ expansion provides the perturbative framework for solving the
equation of motion if higher powers of $Q$ are included.
To next-to-leading order in $1/N_c$ one can write
\beq
Q = \frac{1}{2 \tau} Q_0 + \delta Q
\eeq
and Eq. (\ref{eoml}) leads then to
\beq
\delta Q = - \frac{\lambda}{ 4 \tau^4} Q_0^3
\eeq
modulo higher corrections in $1/N_c$, i.e. higher orders of $Q_0$.
Reinserting the solution for $Q$ into the effective Lagrangian one obtains a
similar Lagrangian as in Eq. (\ref{high}), but with higher orders in 
$Q_0 = \sqrt{6} \eta_0/F_0 + \theta $ which for the sake of brevity are not
shown here.
Therefore, going up to higher powers of $Q$ is similar to expanding the
functions $V_i$ to higher orders in $\sqrt{6} \eta_0/F_0 + \theta $.
Having examined the impact of higher orders of $Q$ in the effective Lagrangian,
we will restrict ourselves to the Lagrangian with factors of $Q$ and $Q^2$
given in Eq. (\ref{q}).

So far we have eliminated the field $Q$ via its equation of motion at lowest
order in the derivatives and quark masses.
We will now proceed by including a term of higher chiral order 
into the equation of motion. In order to keep the arguments as simple
as possible we restrict ourselves to the term 
$i  \langle U^{\dagger} \nabla_{\mu}U \rangle  \partial^{\mu} Q$.
The inclusion of further terms such as $i  Q \langle \chi_- \rangle$ is
straightforward and can be treated in a similar way.
The equation of motion is then derived from the Lagrangian
\beq
\delta {\cal L}  =  \frac{i}{2} Q \langle \log U - \log U^\dagger \rangle 
   - \theta Q + \tau Q^2 
  + i v_5 \langle U^{\dagger} \nabla_{\mu}U \rangle  \partial^{\mu} Q  
\eeq
and reads
\beq
i v_5 \partial_{\mu} \langle U^{\dagger} \nabla^{\mu}U \rangle  =
 - Q_0 + 2 \tau Q .
\eeq
We can decompose the solution for $Q$ into the piece at lowest chiral order
$\frac{1}{2 \tau} Q_0$ and a small perturbation $\Delta Q$
\beq
Q = \frac{1}{2 \tau} Q_0 + \Delta Q
\eeq
so that
\beq
\Delta Q = \frac{i}{2 \tau} v_5 \partial_{\mu} \langle U^{\dagger}
\nabla^{\mu}U \rangle .
\eeq
Inserting $Q$ into the Lagrangian in Eq. (\ref{q}) and neglecting terms of
higher chiral orders, the only additional terms linear in
$ \Delta Q$ read
\beq
\delta {\cal L} = - \Big(\sqrt{6} \eta_0/F_0 + \theta \Big) \Delta Q
  + \tau 2 \frac{Q_0}{2 \tau} \Delta Q = 0 .
\eeq
Therefore, taking only the term $ i v_5 \langle U^{\dagger} \nabla_{\mu}U
\rangle  \partial^{\mu} Q $ into account and working to second order in the
derivative expansion, the additional terms in the Lagrangian happen to cancel.
But in general the procedure of eliminating $Q$ via its equation of motion
perturbatively in the derivative or quark mass expansion will produce terms of
higher chiral orders and will lead to the renormalization of the pertinent
couplings of such terms.
This concludes the proof of the equivalence of the Lagrangian which explicitly
includes the topological charge density with the one given in the last section
up to any order both in the derivative expansion and in $Q$.
At this point, we would like to stress that in order to prove the equivalence,
it is essential that $Q$ is eliminated via its classical equation of
motion. Using the equation of motion with quantum corrections for $Q$ would
destroy the equivalence, since this would lead, e.g., to nonanalytic
expressions in the quark masses which cannot be absorbed by a Lagrangian that
is a polynomial in the quark mass matrix $\cal{M}$. 
Such nonanalytic terms are absent in the approach of \cite{GL,Leu}.
Furthermore, the quantum corrections of the equation of motion for $Q$ are in
general divergent and have to be regularized. This leads  to scale 
dependent contributions which must
be compensated by a suitable redefinition of the pertinent coupling constants.

\section{Inclusion of baryons}
After having ensured ourselves that the approaches discussed above are
equivalent in the purely mesonic sector, we can now proceed by including the
ground state baryon octet in the effective theory.
To this end, it is convenient to summarize the meson fields
in an object of axial-vector type with one derivative
\beq
u_{\mu} = i u^\dagger \nabla_{\mu} U u^\dagger .
\eeq
The matrix $ u_{\mu}$ transforms under $U(3)_R \times U(3)_L$
as a matter field,
\beq
u_{\mu} \rightarrow u_{\mu}' = K u_{\mu} K^\dagger
\eeq
with $K(U,R,L)$ the compensator field representing an element of the conserved
subgroup $ U(3)_V$.
In the context of the first scheme the baryonic Lagrangian up to linear 
order in the derivative expansion has already been
given in \cite{B} and reads
\beqa  \label{bar0}
{\cal L}_{\phi B} &=& i W_1 \langle [D^{\mu},\bar{B}]\gamma_{\mu} B
\rangle - i W_1^* \langle \bar{B}  \gamma_{\mu}  [D^{\mu},B] \rangle 
+ W_2 \langle \bar{B}B \rangle \no \\
&& + W_3 \langle \bar{B} \gamma_{\mu}
 \gamma_5 \{u^{\mu},B\} \rangle  
+  W_4 \langle \bar{B} \gamma_{\mu} \gamma_5 [u^{\mu},B] \rangle 
+ W_5 \langle \bar{B} \gamma_{\mu} \gamma_5 B \rangle \langle u^{\mu} \rangle
\no \\
&& + W_6 \langle \bar{B} \gamma_{\mu} \gamma_5 B \rangle \nabla^{\mu} \theta
+ i W_7 \langle \bar{B} \gamma_5 B \rangle 
\eeqa
with $D_{\mu}$ being the covariant derivative of the baryon fields
and the baryon octet $B$ is given by the matrix
\beqa
 B =  \left(
\matrix { {1\over \sqrt 2} \Sigma^0 + {1 \over \sqrt 6} \Lambda
&\Sigma^+ &p \nonumber \\
\Sigma^-
        & -{1\over \sqrt 2} \Sigma^0 + {1 \over \sqrt 6} \Lambda & n
        \nonumber \\
\Xi^-
        &  \Xi^0 &- {2 \over \sqrt 6} \Lambda  \nonumber \\} 
\!\!\!\!\!\!\!\!\!\!\!\!\!\!\! \right) \, \, \, \, \,  
\end{eqnarray}
which transforms as a matter field
\beq
B \rightarrow B' = K B K^\dagger .
\eeq
The $W_i$ are functions of the combination $\sqrt{6} \eta_0/F_0 + \theta$.
From parity it follows that they are even in  this variable except
$W_7$ which is odd.

If one prefers to include the background field $Q$ explicitly, see
e.g. \cite{Ste}\footnote{In \cite{Ste} a subset of the Lagrangian
considered here has been discussed.}, 
the baryonic Lagrangian reads up to quadratic terms in $Q$
\beqa
{\cal L}_{\phi B} &=& i \Big( - \frac{1}{2} + \alpha Q^2 \Big) 
    \langle [D^{\mu},\bar{B}]\gamma_{\mu} B \rangle 
 - i \Big( - \frac{1}{2} + \alpha^* Q^2 \Big) 
      \langle \bar{B}  \gamma_{\mu}  [D^{\mu},B] \rangle   \no \\
&& + \Big( - \mnod + \beta Q^2 \Big) \langle \bar{B}B \rangle 
  + \Big( -  \frac{1}{2} D + \gamma Q^2 \Big) \langle \bar{B} \gamma_{\mu}
 \gamma_5 \{u^{\mu},B\} \rangle  \no \\
&&  + \Big( -  \frac{1}{2} F + \delta Q^2 \Big)
    \langle \bar{B} \gamma_{\mu} \gamma_5 [u^{\mu},B] \rangle
  + \Big( -  \frac{1}{2} \Lambda + \epsilon Q^2 \Big)  \langle \bar{B} 
   \gamma_{\mu} \gamma_5 B \rangle \langle u^{\mu} \rangle   \no \\
&&  + i \kappa Q   \langle \bar{B} \gamma_5 B \rangle
 + \lambda \langle \bar{B} \gamma_{\mu} \gamma_5 B \rangle \partial^{\mu} Q .
\eeqa
Taking $Q$ from the equation of motion at lowest order as given in 
Eq. (\ref{low}) one obtains
\beqa
{\cal L}_{\phi B} &=& i\Big(-\frac{1}{2} + \frac{\alpha}{4 \tau^2} Q_0^2 \Big) 
    \langle [D^{\mu},\bar{B}]\gamma_{\mu} B \rangle 
 - i \Big( - \frac{1}{2} + \frac{\alpha^*}{4 \tau^2} Q_0^2 \Big) 
      \langle \bar{B}  \gamma_{\mu}  [D^{\mu},B] \rangle   \no \\
&& + \Big( -\mnod +\frac{\beta}{4 \tau^2} Q_0^2 \Big) \langle \bar{B}B \rangle 
  + \Big( -  \frac{1}{2} D + \frac{\gamma}{4 \tau^2} Q_0^2 \Big) 
   \langle \bar{B} \gamma_{\mu} \gamma_5 \{u^{\mu},B\} \rangle  \no \\
&&  + \Big( -  \frac{1}{2} F + \frac{\delta}{4 \tau^2} Q_0^2 \Big)
    \langle \bar{B} \gamma_{\mu} \gamma_5 [u^{\mu},B] \rangle
  + i \frac{\kappa}{2 \tau} Q_0   \langle \bar{B} \gamma_5 B \rangle \no \\
&&  + \Big( -  \frac{1}{2} \Lambda - \frac{\lambda}{2 \tau} 
     + \frac{\epsilon}{4 \tau^2} Q_0^2  \Big)  \langle \bar{B} 
   \gamma_{\mu} \gamma_5 B \rangle \langle u^{\mu} \rangle  
 +  \frac{\lambda}{2 \tau} \langle \bar{B} \gamma_{\mu} 
  \gamma_5 B \rangle \nabla^{\mu}  \theta .
\eeqa
This Lagrangian agrees with the one in Eq. (\ref{bar0}) after expanding the
functions $W_i$ and taking only the lower orders into account. Higher powers of
$Q$ correspond to higher orders in the expansion of the $W_i$.

This time it is of particular interest what kind of
modifications in the Lagrangian
result if $Q$ is eliminated via an equation of motion which 
includes the baryons.
For simplicity we will restrict ourselves to the equation of motion which
results from the Lagrangian
\beq
\delta {\cal L} = - Q_0 Q + \tau Q^2 + i \kappa Q   \langle \bar{B} \gamma_5 B
\rangle .
\eeq
The pertinent equation of motion reads
\beq
- Q_0 + 2 \tau Q + i \kappa  \langle \bar{B} \gamma_5 B \rangle = 0
\eeq
with the solution
\beq
Q = \frac{1}{2 \tau}  \Big( Q_0 - i \kappa  \langle \bar{B} \gamma_5 B 
       \rangle  \Big).
\eeq
Reinserting the solution for Q into the Lagrangian gives rise to terms quartic
in the baryons, e.g. $\langle \bar{B} \gamma_5 B \rangle \langle \bar{B}
\gamma_5 B \rangle $. Since we restricted ourselves from the beginning to the
one-baryon sector, we can drop these additional contributions. But in general
such terms will renormalize the parameters of an effective theory with more
baryons.

\section{Summary}
In this work we have shown the equivalence of two different frameworks which
have been proposed to include the $\eta'$ in chiral perturbation theory both in
the purely mesonic sector and in the presence of the ground state baryon octet.
In the first approach, one starts with an effective chiral Lagrangian which is
invariant under axial $U(1)$ rotations. This is achieved by treating the vacuum
angle $\theta$ as an external field $\theta(x)$ which transforms under $U(1)_A$
in such a way that it compensates the term added to the QCD Lagrangian by the
anomaly.

In the second framework, one keeps the topological charge density $Q$ as a
background field within the effective theory. The chiral Lagrangian includes a
term proportional to $Q$ which is not invariant under $U(1)_A$ and reproduces
the anomaly in the divergence of the axial-vector current.
The field $Q$, on the other hand, is treated as $U(1)_A$ invariant and is
eliminated  via its classical equation of motion. The 
$U(1)_A$ invariance of $Q$ is only fulfilled if one adds a parity violating
piece $\theta Q$ to the Lagrangian and proposes the same transformation law
under $U(1)_A$ for $\theta$ as in the first scheme.
The relation between the different sets of parameters in both Lagrangians is
clarified for arbitrary high powers of $Q$. 
From our discussion it becomes clear, that the first procedure of starting
with an $U(1)_A$ invariant Lagrangian is simpler, although the second framework
might serve as a check for deriving the effective Lagrangian.

\section*{Acknowledgments}
Useful discussions with S. Bass, N. Kaiser, and W. Weise are 
gratefully acknowledged.

\end{document}